\documentstyle[osa,aps,prl,epsfig]{revtex}
\begin{document}
\twocolumn[\hsize\textwidth\columnwidth\hsize\csname
@twocolumnfalse\endcsname
\title{
Observation of the Mott Transition in VO$_2$ Based Transistors}
\author{Hyun-Tak Kim$^{\ast}$, B. G. Chae, D. H. Youn, S. L. Maeng, K. Y. Kang}
\address{Telecom. Basic Research Lab., ETRI, Daejeon 305-350, South Korea}
\date{August 2, 2003}
\maketitle{}
\begin{abstract}
An abrupt Mott metal-insulator transition (MIT) rather than the
continuous Hubbard MIT near a critical on-site Coulomb energy
$U/U_c$=1 is observed for the first time in VO$_2$, a strongly
correlated material, by inducing holes of about 0.018$\%$ into the
conduction band. As a result, a discontinuous jump of the density
of states on the Fermi surface is observed and inhomogeneity
inevitably occurs. The gate effect in fabricated transistors is
clear evidence that the abrupt MIT is induced by the excitation of
holes. \\PACS numbers: 71.27. +a, 71.30.+h
\\
\end{abstract}
]
\newpage
In a strongly correlated system, a metal-insulator transition
(MIT) near a critical on-site Coulomb energy, $U/U_c$=1, has long
been controversial in terms of whether the transition is abrupt or
continuous in experiments [1-4], although a first-order MIT with
temperature was observed by Morin [5]. An abrupt MIT indicates the
Mott transition (first order) and a continuous MIT is the Hubbard
transition (second order). The MIT breaks down an energy gap,
formed by the strongly correlated Coulomb energy, between
sub-bands in a main band. Mott first predicted that the abrupt MIT
occurs when a lattice constant is larger than a critical value
[1]. Brinkman and Rice theoretically demonstrated an abrupt MIT
near $U/U_c$=1 for a strongly correlated metal with an electronic
structure of one electron per atom [6]. Hubbard first derived
that, when sub-bands overlap just below $U_c$, there is a finite
minimum density of states (DOS) at the Fermi level, the DOS
increases with decreasing $U$, and the system is metallic [7];
this is Hubbard's continuous MIT. Later, the continuous MIT was
confirmed in the infinite-dimensional Hubbard model [2].

Applying an electric field to a two-terminal structure, Kumai
$et~al.~[8]$ measured Ohmic behavior in an organic Mott insulator
in a regime, where conduction from nonconduction (insulating
behavior) occurs. Through a theoretical consideration based on the
Hubbard model, Oka $et~al.~[9]$ described the Ohmic behavior in
terms of a universal Landau-Zener quantum tunneling. Boriskov
$et~al.~[10]$ also observed a similar metallic behavior for VO$_2$
to that measured by Kumai $et~al.~[8]$. Thus, on the basis of the
metallic behaviors and Oka's analysis, the MIT just below $U_c$
seems to follow Hubbard's continuous model. However, considering
the abrupt MIT observed in resistance measurement (Fig. 1), an
abrupt MIT in the electric field should also be found.

In this letter, we observe an abrupt jump of current (or DOS) at
an electric field in a two-terminal structure and measure the gate
effect of the jump in a three-terminal device (switching
transistor). The abrupt jump is analyzed in terms of an abrupt MIT
(Mott transition). Inducing internal hole charges of about
0.018$\%$ in hole levels into the conduction band with a
source-drain field or a gate field of a fabricated transistor [11]
is an effective method of revealing the MIT mechanism. Note that
Mott criterion, $n_c~{\approx}~3{\times}10^{18}~cm^{-3}$, is
obtained from $n_c^{1/3}a_H\approx$0.25, where $a_H$ is the Bohr
radius for VO$_2$ [12]. $n_c$ corresponds to about 0.018$\%$ of
the number of carriers in the half-filled band, when one electron
in the cell volume, 59.22$\times$10$^{-24}$ cm$^3$, of VO$_2$ is
assumed; the number of electrons is about
1.7$\times$10$^{22}$~cm$^{-2}$.

\begin{figure}
\vspace{-0.2cm}
\centerline{\epsfysize=9cm\epsfxsize=8.4cm\epsfbox{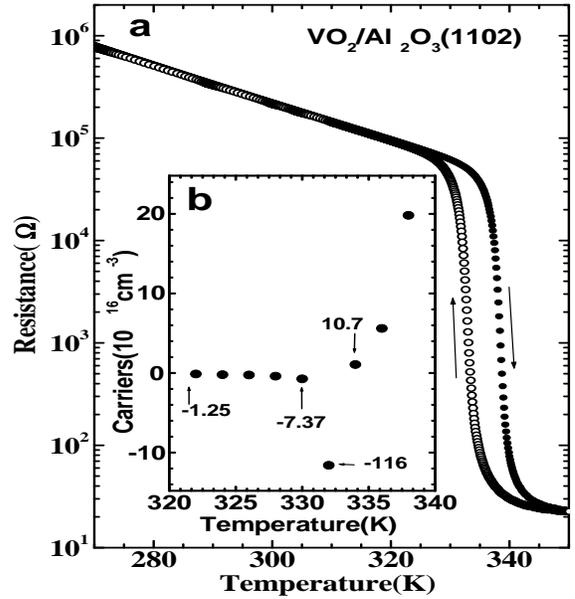}}
\vspace{0.3cm} \caption{${\bf a}$, Temperature dependence of the
resistance of a VO$_2$ film. Hysteresis is shown. ${\bf b}$, The
number of carriers measured by Hall effect. A change of carriers
from holes to electrons is shown at 332 K. The minus sign
indicates that the carriers are holes.}
\end{figure}

Thin films of a Mott insulator, VO$_2$, with a sub-energy gap of
about 1~$eV$ in the $d$ band [13,14], have been deposited on
Al$_2$O$_3$ and Si substrates by laser ablation. The thickness of
the VO$_2$ films is about 900$\AA$. The resistance of the film
decreases with increasing temperature and shows an abrupt MIT at a
transition temperature, $T_{tr}$=340 K (68$^{\circ}$C) (Fig. 1a).
This corresponds with that measured by De Natale $et~al.$ [15] and
Borek $et~al.$ [16]. It was proposed that the abrupt MIT is the
structural phase transition from monoclinic below $T_{tr}$ to
tetragonal above $T_{tr}$ [5]. The decrease of the resistance up
to 340 K indicates an increase of hole carriers, and two kinds of
electron and hole carriers coexist near $T_{tr}$=340 K (Fig. 1b).
From 332 to 340 K, the number of carriers is not discernable
because of mixing of electrons and holes. We speculate that the
number of hole carriers may be the Mott criterion,
$n_c~{\approx}~3{\times}10^{18}~cm^{-3}$, at $T_{tr}$=340 K on the
general basis that an exponential decrease of the resistance with
temperature in semiconductor physics indicates an exponential
increase of carriers. Generally, in oxide materials, there are
holes of about 5.5${\times}$10$^{18}cm^{-3}$ which corresponds to
0.034$\%$ to $d$-band charges $[17-19]$. The holes are coupled to
optical phonons [17]. In the metal regime above 340 K, the
carriers are electrons (Fig. 1).

\begin{figure}
\vspace{-1.1cm}
\centerline{\epsfysize=7cm\epsfxsize=8.0cm\epsfbox{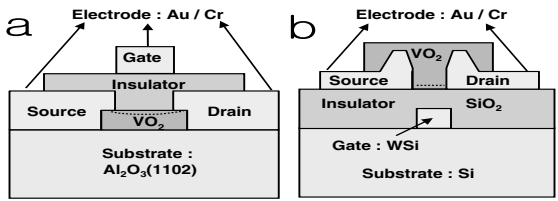}}
\vspace{-2.5cm} \caption{Schematic diagram of transistors. The
dotted line between VO$_2$ and insulator is a channel. The gate
insulators are an amorphous Ba$_{0.5}$Sr$_{0.5}$TiO$_3$ for
transistor 1 (Fig. a), and thermally treated amorphous SiO$_2$ for
transistors 2 and 3 (Fig. b).}
\end{figure}

We fabricated transistors to observe the Mott transition. A
schematic diagram of the transistors is shown in Fig. 2.
Transistor 1 of a channel length, $L_{ch}$= 3${\mu}m$, and a gate
width, $L_w=50{\mu}m$, was fabricated on Al$_2$O$_3$ substrates by
lithography processes (Fig. 2a). A gate insulator of transistor 1
was an amorphous Ba$_{0.5}$Sr$_{0.5}$TiO$_3$ (BSTO) which was
deposited on the VO$_2$ films. The interface between the VO$_2$
film and the amorphous gate insulator was sharp. A gate-source
current, $I_{GS}$, between the gate and the source for transistor
1 is an order of 10$^{-13}~A$, which indicates that there is
sufficient insulation between the gate and the source. Transistors
2 and 3 of a gate length, $L_{ch}=5{\mu}m$, and a gate width,
$L_w=25{\mu}m$, were manufactured on Si substrates. Their
structure is shown in Fig. 2b. SiO$_2$ as the gate insulator was
thermally treated. It is revealed that an interface between the
polycrystal VO$_2$ film and the amorphous SiO$_2$ film is not
sharp and complicated, and that the VO$_2$ films are inhomogeneous
[20]. However, the SiO$_2$ insulator is strong with respect to a
high field and is superior to the BSTO insulator for electronic
application. Au/Cr electrodes were prepared for Ohmic contact. WSi
is used as gate electrode. Characteristics of the transistors were
measured by a precision semiconductor parameter analyzer
(HP4156B). To protect transistors from excess current, the maximum
current was limited to 20mA.

\begin{figure}
\vspace{-0.0cm}
\centerline{\epsfysize=10cm\epsfxsize=8.4cm\epsfbox{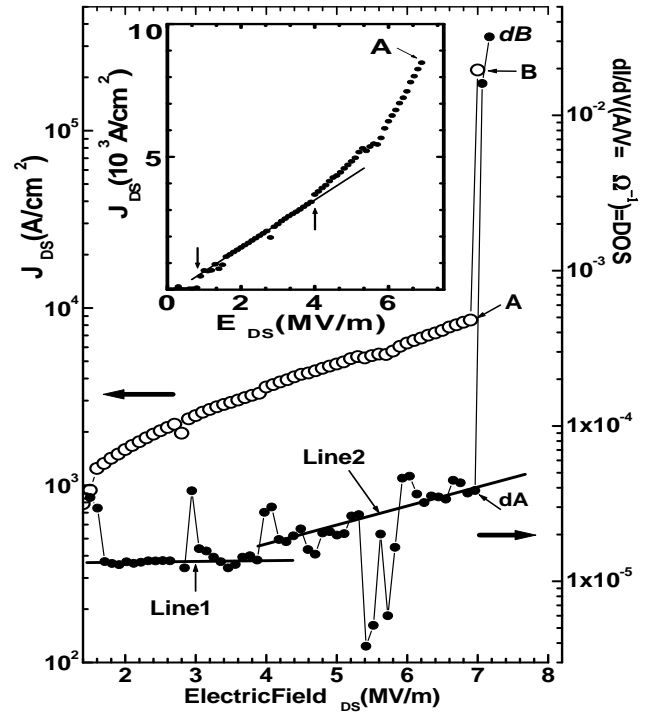}}
\vspace{0.5cm} \caption{Left-axis side: $J_{DS}$ vs. $E_{DS}$ for
transistor 1. Right-axis side: Density of states (DOS), $dI/dV$,
on the Fermi surface obtained from derivation with respect to
$V_{DS}$. Inset: Ohmic behavior from $E_{DS}$=0.8 up to 4 MV/m is
shown. The maximum current was limited to 20mA to protect the
transistor.}
\end{figure}

Figure 3 shows the drain-source current density, $J_{DS}$, vs. the
drain-source electric field, $E_{DS}$, for transistor 1, measured
at $V_{gate}$=0 V (two-terminal structure). $J_{DS}$ and $E_{DS}$
are obtained from the drain-source current and the drain-source
voltage, respectively; $I_{DS}$=$J_{DS}S_{DS}$($S_{DS}$: cross
section) and $V_{DS}=E_{DS}L_{ch}$. $J_{DS}$ behavior below point
A (Inset of Fig. 3), as observed by Boriskov $et~al.~[10]$ and
Kumai $et~al.~[8]$ who used an organic Mott-insulator, is linear
from $E_{DS}\approx$0.8 to 4MV/m, but is nonlinear in the total
regime. It was suggested that the linear Ohmic behavior occurs due
to an applied field $[10]$ and an induced current [8], not an
increase of sample temperature due to leakage current. The Ohmic
behavior was well described through a theoretical consideration in
terms of a universal Landau-Zener quantum tunneling based on the
Hubbard model [9]. However, since holes were observed below 334 K
(Fig. 1b), it is asserted that the carriers for the Ohmic behavior
are holes and the number of holes is very small. Thus, the Ohmic
behavior is not an intrinsic property of metal and may be due to
scattering of only a few carriers existing at room temperature by
a weak electric field because the density of states (DOS) does not
depend on the electric field (line 1 in Fig. 3). Moreover, the
nonlinear behavior is semiconduction due to the increase of hole
carriers by Zener's impact ionization and is a doping process
wherein $n_c$ of holes are induced by the electric field. This is
supported by the fact that the DOS increases exponentially with an
increasing electric field (line 2 of Fig. 3). Thus, the VO$_2$
film below point A is regarded as a semiconductor.

\begin{figure}
\vspace{0.0cm}
\centerline{\epsfysize=15.2cm\epsfxsize=8.6cm\epsfbox{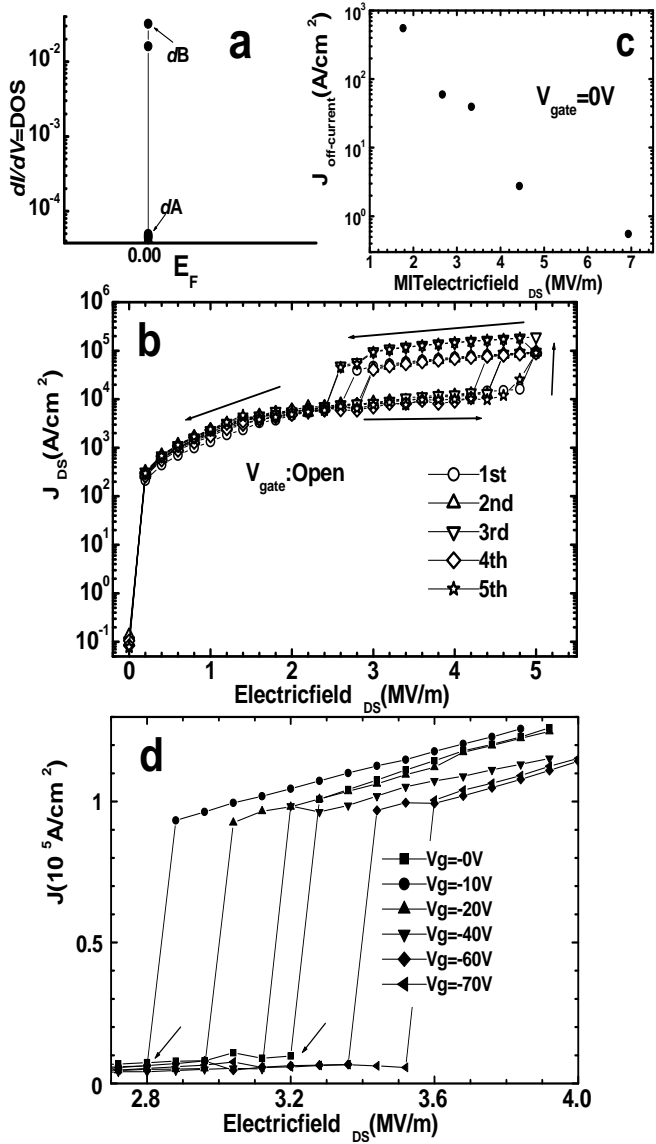}}
\vspace{0.5cm} \caption{${\bf a}$, DOSs in Fig.3 are redrawn with
respect to the Fermi level. E$_F$ denotes the Fermi level. ${\bf
b}$, Hysteresis loops measured in transistor 2. Measurements of 5
times were continuously carried out. ${\bf c}$, Off-current vs.
MIT-$E_{DS}$. The off-currents (or leakage current) were extracted
from 5 transistors. ${\bf d}$, $J_{DS}$ vs. $E_{DS}$ near the
abrupt jump of transistor 3. MIT electric field increases with
increasing negative gate voltage from $V_g$=-10V. Above
$J_{DS}{\approx}$0.9$\times$10$^5$~A/cm$^2$, Ohmic behavior is
exhibited.}
\end{figure}

Abrupt jumps of $J_{DS}$ at point A and the DOS at point $dA$ are
shown in Fig. 3. The measured maximum-current density at point B
is $J_{measured}\approx$~2$\times$10$^5$ A/cm$^2$, which is an
order of the current density observable in a dirty metal. Note
that the true maximum current density is much higher than
$J_{measured}$, because the measurement was limited to 20mA. The
jump of $J_{DS}$ between point A and point B corresponds to a jump
of conductivity, $\sigma$, because $J_{DS}$ is
${\sigma}E=n_{free}ev$ and $E$ is constant between points A and B;
${\sigma\propto}n_{free}$, where $n_{free}$ is the number of
electron carriers (above point A) on the Fermi surface, and $v$ is
the carrier velocity on the Fermi surface and is constant between
points A and B due to constant $E$. This indicates that the number
of carriers discontinuously increases, and that the DOS on the
Fermi surface jumps from points dA to dB (Fig. 3). The jump of the
DOS is a typical behavior of a first-order MIT which was first
theoretically derived by Brinkman and Rice [6]. Note that carriers
for current measured in semiconductors and metals are on the Fermi
surface. Derivatives, $dI/dV$, at respective fields correspond to
the DOSs on the Fermi surface (Fig. 3). The DOS in Fig. 3 can be
expressed in terms of the Fermi level (Fig. 4a). Moreover,
hysteresis loops of 5 times for transistor 2 were continuously
measured (Fig. 4b), which may be due to the Joule heating and is
also evidence of a first-order MIT. The abrupt jump was also
measured more than 1,500 times without breakdown in a transistor.
Fig. 4d shows Ohmic behavior after jump, which indicates that this
phase is metal. Thus, we suggest that the jump at point A is the
abrupt MIT (or Mott transition). If Hubbard's continuous MIT
exists, the jump should be not observed and the non-linear
behavior with an increasing field in the inset of Fig. 3 should be
continuously exhibited from point A to point B in the electron
system. However, continuous behavior is not found.

Figure 4c shows the off-current density, $J_{off-current}$, vs.
the MIT drain-source electric field, MIT-$E_{DS}$, at $V_{gate}$=0
$V$ (two-terminal structure). Although $V_{gate}$=0 $V$, the
gate-source currents were very small to be ignored. The
off-current is defined as $I_{DS}$ near $V_{DS}$=0V and $V_g$=0V
(or gate open); other researchers refer to off-current as
leakage-current. Data were selected from 5 transistors.
MIT-$E_{DS}$ increases with decreasing $J_{off-current}$, which
also provides information for revealing the mechanism of the
abrupt jump. The off-current is caused by the excitation of holes
in impurity levels such as oxygen deficiency. When the number of
total holes in the hole levels is given by $n_{tot}=n_b +
n_{free}$, where $n_b$ is the number of bound holes in the levels
and $n_{free}$ is the number of holes freed from the levels. $n_b$
decreases with increasing $n_{free}$, because $n_{tot}$ is
constant. The larger off-current is attributed to the increase of
$n_{free}$. For the abrupt jump,
${\triangle}n{\equiv}n_c-n_{free}$=0 should be satisfied, where
$n_c~{\approx}~3{\times}10^{18}~cm^{-3}$, as predicted by Mott.
Hence, the decrease of ${\triangle}n$ (increase of $n_{free}$)
contributes to the reduction of the MIT-$E_{DS}$.

Figure 4d shows $J_{DS}$ vs. $E_{DS}$ near the abrupt jump of
transistor 3. Its characteristics are as follows. First, the gate
effect at $E_{DS}$=2.8MV/m and $V_{gate}$=-10V is due to induced
holes and occurs suddenly; this indicates attainment of $n_c$.
Second, the MIT-$E_{DS}$ increases with increasing negative gate
voltage (or field), which is due to a decrease of
 the conductivity; $J_{DS}$s at MIT points can be nearly regarded as
constant and $\sigma_{DS}=J_{DS}/E_{DS}$. This is due to an
increase of hole carriers generated by the negative gate fields
and indicates an increase of inhomogeneity (injection of holes to
electron system). It was also observed that the MIT-$E_{DS}$
decreases with increasing positive gate voltage when the
off-current is large. Third, in the metal (electron system) regime
over $J_{DS}$=0.9$\times$10$^4$ mA/cm$^2$, Ohmic behavior differs
from the Ohmic behavior in the hole system in Fig. 3 and arises
from a dirty interface between the polycrystal VO$_2$ film and the
amorphous SiO$_2$. The dirty interface causes resistance [20] and
is a channel where current flows (Fig. 2b). We suggest that the
Ohmic behavior in Fig. 4d is a true metallic characteristic.
Finally, the gate induced abrupt MIT reveals that the MIT depends
upon the hole carriers in the semiconductor regime, and that
inhomogeneity also inevitably arises from hole doping [21].

 In conclusion, in the Mott insulator of VO$_2$,
 MIT near $U/U_c$=1 abruptly occurs in company with inhomogeneity
 through semiconduction as a doping
process of internal holes of $n_c{\approx}0.018\%$. This will be
observed in all Mott insulators.

We thank Dr. Soo-Hyeon Park at KBSI for Hall-effect measurement,
Dr. Gyungock Kim for valuable discussions on the Zener transition,
and Dr. J. H. Park for fabrication of Si$_3$N$_4$ films with CVD.
HT Kim, the leader of this project, developed the concept, and
wrote the paper. BG Chae and DH Youn deposited VO$_2$ and BSTO
films, performed the transistor fabrication process including Si
based transistors, and measured $I-V$ characteristics. KY Kang
prepared the laser-ablation and lithography equipments and
generated this project with HT Kim. SL Maeng evaluated transistor
characteristics, the shielding measurement system, and the
Si$_3$N$_4$ film fabrication.

\end{document}